\documentclass[twocolumn]{aastex61}

\usepackage{natbib,aas_macros} 
\usepackage{hyperref}
\usepackage{graphicx}	
\usepackage{amsmath,amssymb} 
\usepackage{xspace}
\usepackage{enumerate}
\usepackage{fp} 
\usepackage{lineno}
\usepackage{subfigure}

\newcommand{\ic}{IC\,1613\xspace}

\shorttitle{Calibration of the \ic NIR-TRGB}
\shortauthors{Madore, Freedman, Hatt et al.}

\begin{document}

\title{The Near-Infrared Tip of the Red Giant Branch. I. \\A Calibration in the Isolated Dwarf Galaxy \ic}

\author{Barry~F.~Madore}\affil{Department of Astronomy \& Astrophysics, University of Chicago, 5640 South Ellis Avenue, Chicago, IL 60637}\affil{Observatories of the Carnegie Institution for Science, 813 Santa Barbara St., Pasadena, CA~91101}\email{barry.f.madore@gmail.com}

\author{Wendy~L.~Freedman}\affil{Department of Astronomy \& Astrophysics, University of Chicago, 5640 South Ellis Avenue, Chicago, IL 60637}

\author{Dylan~Hatt}\affil{Department of Astronomy \& Astrophysics, University of Chicago, 5640 South Ellis Avenue, Chicago, IL 60637}

\author{Taylor~J.~Hoyt}\affil{Department of Astronomy \& Astrophysics, University of Chicago, 5640 South Ellis Avenue, Chicago, IL 60637}

\author{Andrew~J.~Monson}\affil{Department of Astronomy \& Astrophysics, Pennsylvania State University, 525 Davey Lab, University Park, PA 16802}

\author{Rachael~L.~Beaton}\altaffiliation{Hubble Fellow}\altaffiliation{Carnegie-Princeton Fellow}\affil{Department of Astrophysical Sciences, Princeton University, 4 Ivy Lane, Princeton, NJ~08544}

\author{Jeffrey~A.~Rich}\affil{Observatories of the Carnegie Institution for Science, 813 Santa Barbara St., Pasadena, CA~91101}

\author{In Sung~Jang}\affil{Leibniz-Institut fur Astrophysik Potsdam (AIP), An der Sternwarte 16, D-14482, Potsdam, Germany}

\author{ Myung~Gyoon~Lee}\affil{Department of Physics \& Astronomy, Seoul National University, Gwanak-gu, Seoul 151-742, Korea}

\author{Victoria~Scowcroft}\affil{Department of Physics, University of Bath, Claverton Down, Bath BA2 7AY, UK}

\author{Mark~Seibert}\affil{Observatories of the Carnegie Institution for Science, 813 Santa Barbara St., Pasadena, CA~91101}

\begin{abstract}

Based on observations from the \emph{FourStar} near-infrared camera on the 6.5m Baade-Magellan telescope at Las Campanas, Chile, we present calibrations of the  $JHK$ 
luminosities of stars defining 
the tip of the red giant branch (TRGB) 
in the halo of the Local Group dwarf galaxy \ic. We employ metallicity-independent (rectified) T-band magnitudes---constructed
using $J,H$ and $K$-band magnitudes and both $(J-H)~ \& ~(J-K)$ colors
in order to flatten the upward-sloping red giant branch tips as otherwise seen in their apparent color-magnitude diagrams. We describe and quantify the advantages of working at these particular near-infrared wavelengths, which are applicable to both \emph{HST} and \emph{JWST}. We also note that these same wavelengths can be accessed from the ground for an eventual tie-in to \emph{Gaia} for absolute astrometry and parallaxes to calibrate the intrinsic luminosity of the TRGB. Adopting the color terms derived from the \ic data, as well as the zero-points from a companion study of the Large Magellanic Cloud whose distance is anchored to the geometric distances of detached eclipsing binaries, 
we find a true distance modulus of 24.32 $\pm$ 0.02~ (statistical) $\pm$ 0.06~mag (systematic) for
\ic, which compares favorably with the recently published multi-wavelength, multi-method consensus modulus of 24.30 $\pm$ 0.05~mag by \citet{hatt17}. 

\end{abstract}


\section{Introduction}

In recent years, the tip of the red giant branch (TRGB) has emerged as a
superb method for measuring the distances to nearby galaxies \citep[e.g.,][]{lee93,sak04,riz07,mag08,jl17}.
This feature, i.e. the sudden truncation of the red giant branch in luminosity, is explained by stellar astrophysics: theoretically, it is predicted that the upward evolution of the luminosities of low-mass stars
\citep[$\lesssim 1.6$~M$_\odot$;][]{gir00} 
on the red giant branch will have a sharp termination as the temperature of the helium core reaches $\sim 10^8$~K, resulting in a lifting of the degeneracy of the core, and allowing stable helium burning to occur as the star rapidly decreases in luminosity, settling onto the horizontal branch \citep[e.g.,][]{iben84,sal97}. TRGB distances have been measured in practice by applying edge-detection techniques to the color-magnitude diagrams of galaxies to mark the transition point between the foreground and the red giant branch, one method being the Sobel filter \citep[for a review see][]{riz07}; or, alternatively fitting a model to the observed luminosity function \citep[e.g.,][]{men02,conn11}.

\begin{figure*}[ht]
\centering 
\includegraphics[width=12.0cm, angle=-90]{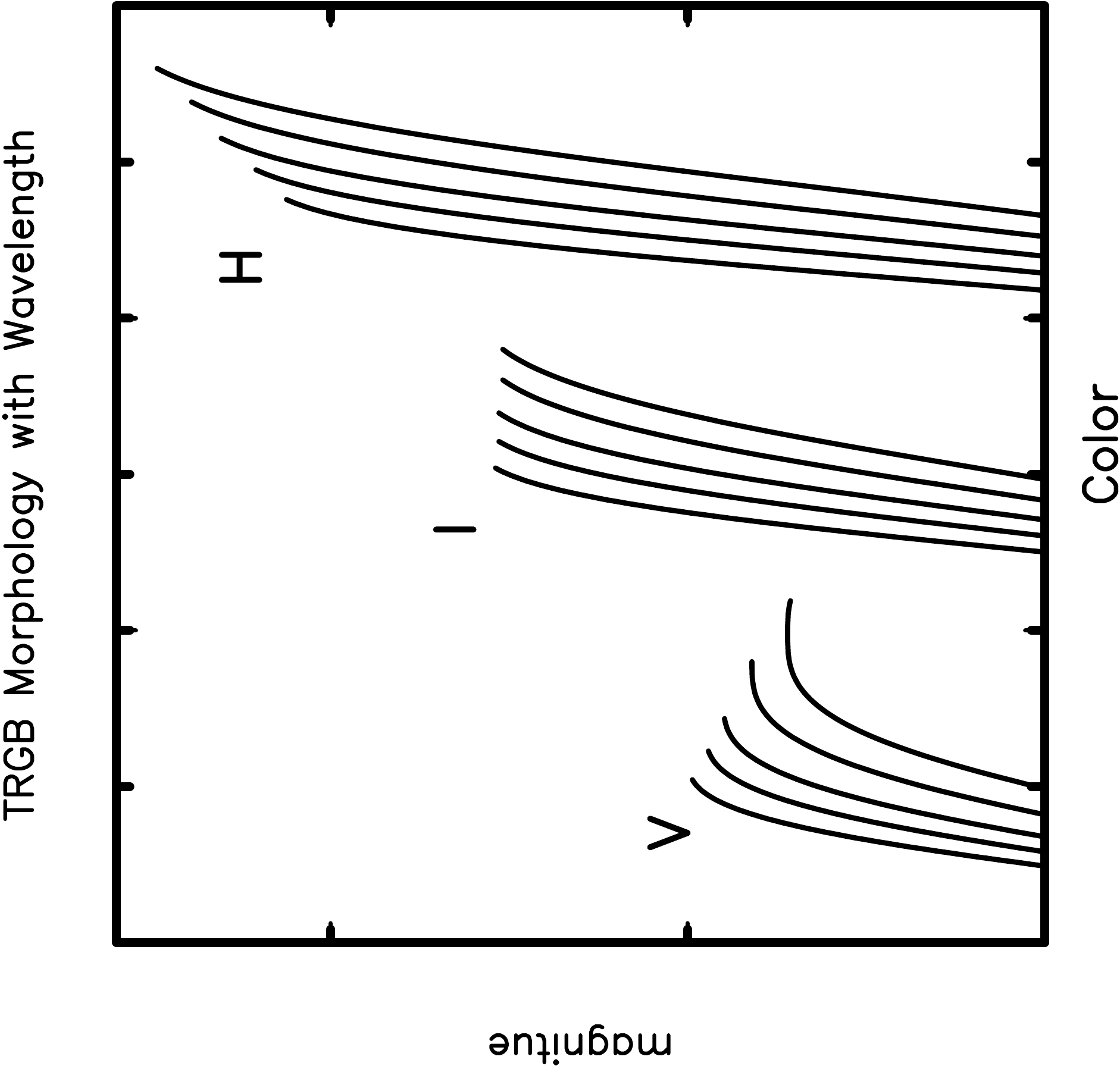}
\caption{Schematic representation of the absolute magnitude and color variation of the red giant branch tip morphology as a function of wavelength.  The left set of progressively downward-sloping branches are representative of $V$-band color-magnitude diagrams. The middle set, still fanning out in color but terminating at approximately the same absolute magnitude, closely approximate the observed behavior in the $I$-band. Finally, the far right set of branches show the upward sloping run of the TRGB as a function of color, typical of $J, H~ \& ~K$-band CMDs. The relative downward- or upward- sloping at the TRGB between branches is due to the radiative-transfer effects of increased metallicity (namely in stellar atmospheres) of the 3 illustrated bands. 
\label{fig:TRGB-diag}} 
\end{figure*}

It is established that the bolometric luminosity at the so-called ``He flash'' is weakly dependent on metallicity at or below solar values \citep[see Figure 5.19 of][]{salaris_stellar_pop}, but there are wavelength-dependent observational effects based on the metal content of stellar atmospheres. In the optical (e.g., $B, V$), the effects of line blanketing suppress the emergent blackbody radiation, which translates into a downward sloping TRGB as redder, higher metallicity stars appear progressively fainter relative to their metal-poor counterparts. Since this radiation inevitably re-emerges as continuum flux in the near- and mid-infrared, the bolometric corrections in the near-infrared (NIR) reverse their sign so that TRGB stars become brighter (both in zero-point and as a function of color). Theoretical studies predict that bolometric corrections in the $I$-band ($\sim 8,000$ \AA) will largely compensate for the effects of metallicity \citep{sal98} such that the $I$-band luminosity is largely independent of color and metallicity. The positions of the TRGB in these three wavelength regimes (visual, red and near-infrared) are schematically illustrated in \autoref{fig:TRGB-diag}. 

The ``flatness'' of the $I$-band TRGB has been borne out empirically \citep[starting with][]{lee93}, making it the standard passband for TRGB studies. Despite the success and simplicity of the $I$-band TRGB as a distance indicator,  there are two major advantages to extending the TRGB technique to NIR passbands: first, the effects of reddening are much diminished over optical wavelengths \citep[e.g., $A_H = 0.18 \times A_V$ vs $A_I = 0.48 \times A_V$;][]{rie85}; and second, since the bolometric corrections in the NIR reverse their sign so that TRGB stars become brighter, they are more easily observable out to greater distances than at optical wavelengths. The near-infrared, however, has not been as well-tested or explored for distance determinations as has the optical.

In addition to the early work of \citet{2007AA...467.1025G}, \citet{bel08} and \citet{2009MNRAS.394..795W}, two more recent studies by \citet{dal12} and \citet{wu14} have explored the use of the NIR-TRGB as an extragalactic distance indicator. Those studies focused on the {\it maximum} luminosity of the tip, however, and (with the exception of \citet{2009MNRAS.394..795W}, who did make insightful comments about the non-zero slope of the $K$-band TRGB) did not fully address the detailed form and/or systematics of the distribution of tip stars in multi-metallicity old stellar populations{ \footnote{ An additional parameter worthy of comment is the mass range of stars contributing to the TRGB. \citep[][their Figure 7.4]{Old_Stellar_Pop} show that for ages between 8 and 14 Gyr there is virtually no sensitivity of the $I$-band tip magnitude to age. And even more recently \citet{2017RNAAS...1...31L} have shown that the tip magnitude in the $J$ band is also unchanged at the 0.005~mag level, for an even larger age range of 5 to 13~Gyr. Nevertheless, we would still argue that it is dangerous to attempt to measure the TRGB in disk fields made up composite stellar populations overlapping the tip and co-existing with interstellar dust intrinsic to those local lines of sight.}, specifically, the strong correlation of TRGB magnitude with color at NIR wavelengths. }

This paper is the first in a series aimed at
a NIR calibration of the detailed structure of the NIR-TRGB as a function of
metallicity. Here, we begin our iterative calibration with the nearby Local Group dwarf irregular galaxy
\ic, which is an isolated and low-metallicity galaxy at a distance of about 725~kpc. Its
proximity, low foreground and internal extinction, and the existence of numerous
independent distance measurements in the literature, all make \ic an ideal target for such a calibration. Moreover, in a companion paper (Hoyt et al., submitted) we provide an absolute calibration of the NIR-TRGB using the Large Magellanic Cloud (LMC), which we apply to this work to obtain a NIR-TRGB distance to \ic.

It is worth noting that over time the calibration of the TRGB zero-point, in the $I$-band, say, has been remarkably stable at the $\pm0.05$~mag level \citep[e.g.,][]{dac90,lee93,riz07,jang17} despite cautionary estimates of the systematic uncertainty to the contrary. 
After decades of gradual progress in measuring the structure and distance scale of the TRGB, optical and NIR, the accuracy and precision of the method are now improving at a considerable pace. High signal-to-noise observations of local galaxies---e.g., this study of \ic and the companion study of the LMC by Hoyt et al.---are allowing for unprecedented resolution of TRGB structure to robustly extend the method into the NIR. Beyond individual studies of galaxies, in the near future \emph{Gaia} will also provide accurate and precise trigonometric parallaxes for thousands of red giant branch stars in the Milky Way that will independently set the TRGB distance scale in multi-wavelength regimes. These substantially more accurate distance scales will coincide with the launch of \emph{JWST}, allowing the application of the NIR-TRGB to distances that are over 3 times greater than currently possible with the $I$-band TRGB, thereby allowing a significant increase in the number of accurate distances for the calibration of Type~Ia supernovae that are used in direct measures of $H_0$.

\section{Observations 
and the Multi-Wavelength Sample
}\label{sec:data}

\begin{figure*}[ht]
\centering 
\includegraphics[width=18.0cm, angle=0]{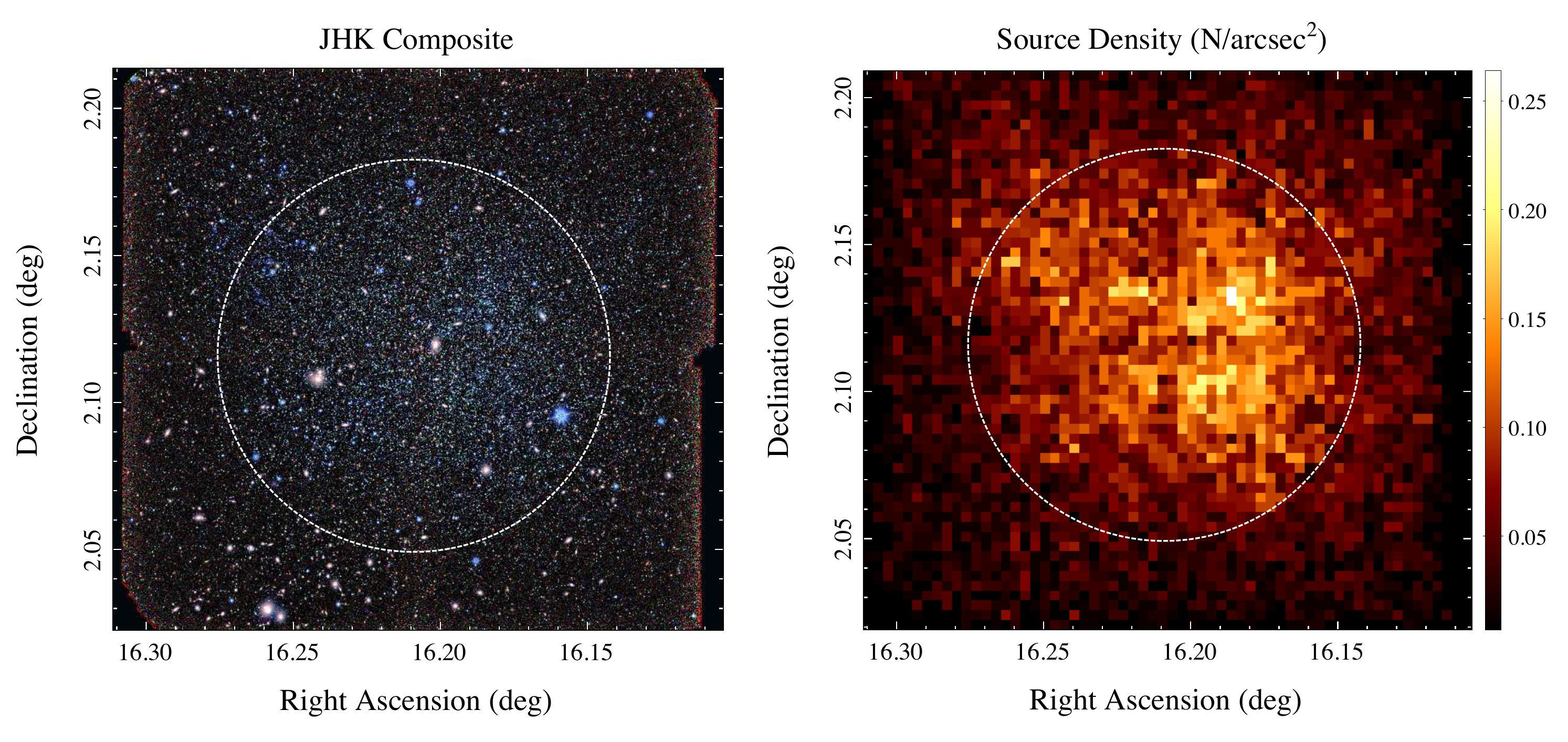} 
\caption{Spatial distribution of stars in \ic contained in the merged $JHK$
catalog. The field of view is $10 \farcm 8 \times 10 \farcm 8$. The inscribed circle [$4\farcm0$ in radius, centered at RA(2000) = 16\fdg2090 and Dec(2000) = +2\fdg1159] shows the inner disk region that has been excluded from the TRGB analysis due to the higher source density and risk of crowding. At right is the detected source density, which is dominated by red giant branch stars, as seen in the color-magnitude diagram presented in Figure~4.}

\label{fig:IC1613xy} 
\end{figure*}

The near-infrared $JHK$ imaging datasets used here are described in detail in \citet{sco13}. Here we note that the \emph{FourStar} imager has field of view of $10 \farcm 8 \times 10 \farcm 8$ and a resolution of 0.159$\arcsec~$pixel$^{-1}$, and that the observations were centered on the main body of \ic so as to optimize detection of Population~I (disk) Cepheids. This field of view is still large enough to cover a significant portion of \ic's outer disk, as is demonstrated in \autoref{fig:IC1613xy}.

Figure 3 shows a plot of signal-to-noise (S/N) ratio in each of the three (JHK) bands as a function of magnitude. Of particular interest in the context of this paper is the S/N ratio obtained at the level of the TRGB. These values are identified in the plot, and range from 70:1 ($\pm$0.014~mag) in the K band, to 120:1 ($\pm$0.009~mag) in the H band 130:1 ($\pm$0.008~mag) in the J band.

\begin{figure*} 
\centering 
\includegraphics[width=9cm,angle=0]{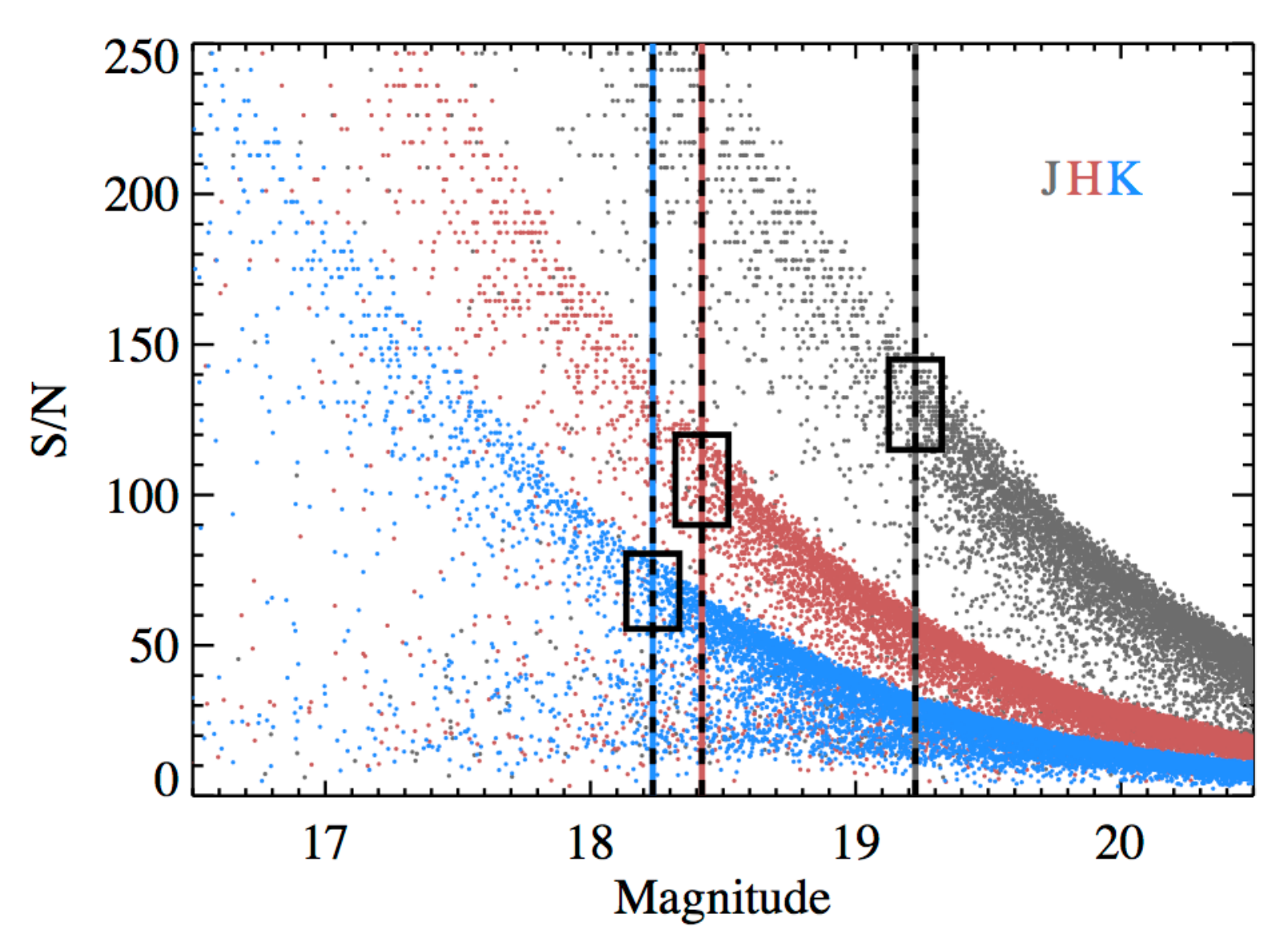} 
\caption{The run of signal-to-noise (S/N) as a function of magnitude for our near-infrared data (J - black, H - red, K - blue). Vertical lines mark the approximate magnitude at which the TRGB is measured, giving values for the  S/N = 130, 110 and 70 for J, H and K, respectively.
\label{fig:IC1613_S2N-small}} 
\end{figure*}

\begin{figure*}[ht]
\centering 
\includegraphics[width=17.0cm, angle=-90]{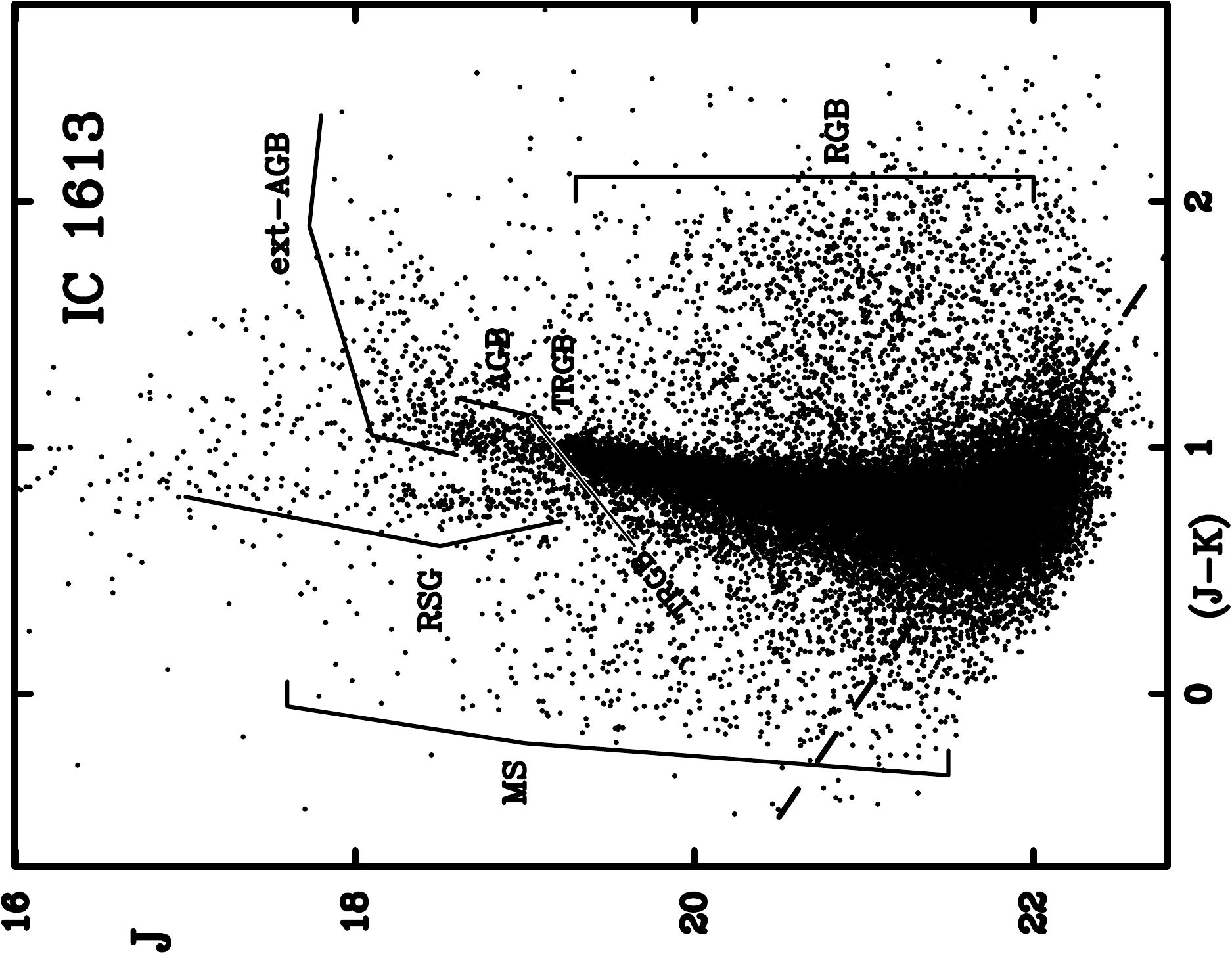} 
\caption{\small The  full, near-infrared $J$ vs $(J-K)$
color-magnitude diagram for the  $10 \farcm 8 \times 10 \farcm 8$  \ic field of view containing 23,525 stars.
Major features are labeled, including the Asymptotic Giant Branch (AGB), Red Supergiants (RSG), extended-AGB stars (ext-AGB), and Red Giant Branch (RGB). The shallow, upward sloping line in the middle of the diagram marks the termination of the red giant branch luminosity function, with AGB stars above the discontinuity and first-ascent RGB stars below. The downward-slanting broken line shows the trace of a $K$-band limiting magnitude of 21.
\label{fig:IC1613-JJK-CMD-2-17}}
\end{figure*}

\autoref{fig:IC1613-JJK-CMD-2-17} shows the $J$ vs $(J-K)$ color-magnitude diagram (CMD) for all 23,525 sources having $JHK$ photometry down to a $J$-band magnitude of about 22~mag. As is immediately apparent at these wavelengths, hot and blue stars, which usually prominently define the high-mass main sequence (MS) in optical CMDs, are still present, but their dominance in optical CMDs is much suppressed in the NIR. The blue limit of the main sequence is broadly outlined by the upward-sloping line to the left of \autoref{fig:IC1613-JJK-CMD-2-17} at a color of $(J-K) \sim$ 0.0~mag. These stars evolve up in luminosity and redward in color, eventually slowing in their evolution at the red supergiant (RSG) sequence, also vertically marked in \autoref{fig:IC1613-JJK-CMD-2-17} at $(J-K) \sim 0.8$~mag. These stars make up the most prominent features in the NIR CMD, attributable to Population~I stars.

\begin{figure*}[ht]
\centering 
\includegraphics[width=17.0cm,angle=-90]{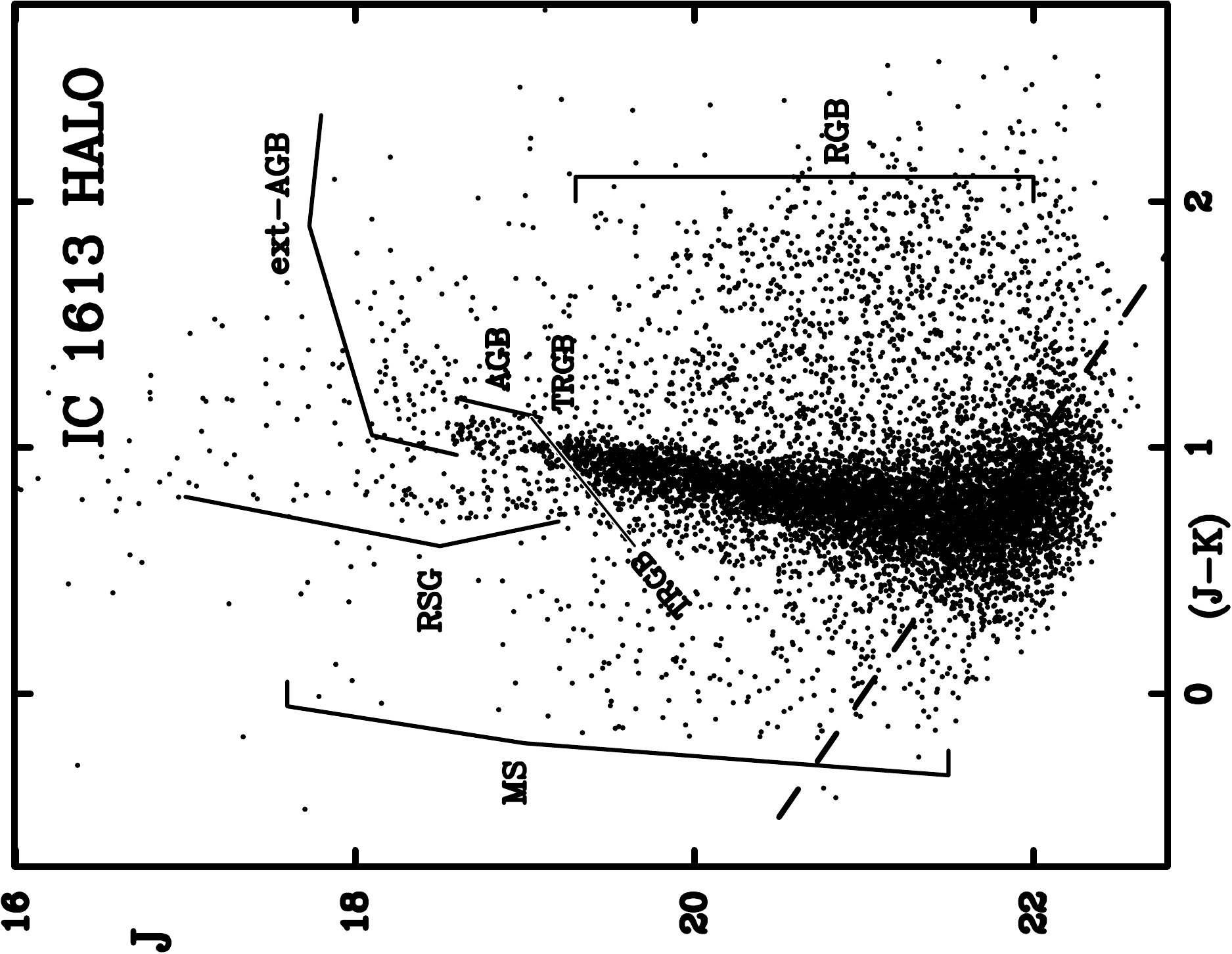} 
\caption{\small The full, near-infrared $J$
vs $(J-K)$ color-magnitude diagram for 6,070 stars in the halo of \ic, restricted here to photometric data with a signal-to-noise of 10. Major features are the same as described in the caption to \autoref{fig:IC1613-JJK-CMD-2-17}. 
\label{fig:IC1613-JJK-CMD-HALO-4-17-mark}} 
\end{figure*}

The remainder of the CMD in \autoref{fig:IC1613-JJK-CMD-2-17} is dominated by Population~II stars: the red giant branch (RGB), the asymptotic giant branch (AGB) and the extended-AGB. The RGB is the dominant feature: its stars rise up from the (significantly fainter) main sequence turn-off of this old population (well below the plotted data) and gradually move to the red by a few tenths of a magnitude in $(J-K)$ color and by nearly 2 mag in luminosity. The dramatic drop in star counts in the CMD above the termination of the RGB (at about $J =$ 19~mag) marks the beginning of the more-rapidly-evolving AGB. This phase continues upward in luminosity for about a magnitude, before it seamlessly merges with the extended-AGB, which quickly fans out to both higher luminosities ($J \sim$ 18~mag) and redder colors, $(J-K) > 2$~mag. For the remainder of this paper, our attention will be drawn to the central portion of this CMD where we will focus on the apparent discontinuity in the RGB luminosity function which corresponds to the termination of the upward luminosity evolution of the first-ascent red giant branch population, i.e. the TRGB.

\begin{deluxetable*}{rrrrrrrcccccc} 
\tabletypesize{\small} 
\tablewidth{0pt} 
\tablecaption{\label{tbl:tbl1}} 
\tablehead{ 
\colhead{ID} &
\colhead{ } &
\colhead{$\alpha$} &
\colhead{ } &
\colhead{ } &
\colhead{$\delta$} &
\colhead{ } &
\colhead{$J$} &
\colhead{$\sigma$} &
\colhead{$H$} &
\colhead{$\sigma$} &
\colhead{$K$} &
\colhead{$\sigma$} \vspace{-0.2cm} \\
\colhead{} &
\colhead{h} &
\colhead{m} &
\colhead{s} &
\colhead{d} &
\colhead{m} &
\colhead{s} &
\colhead{mag} &
\colhead{mag} &
\colhead{mag} &
\colhead{mag} &
\colhead{mag} &
\colhead{mag} 
}
\startdata 
  01603 & 01 & 04 & 59.93 & +02 & 01 & 17.86 &  18.77 &   0.03 &  18.21 &   0.03 &  17.69 &   0.03\\
  29036 & 01 & 04 & 51.30 & +02 & 01 & 18.04 &  21.72 &   0.11 &  20.96 &   0.17 &  20.46 &   0.20\\
  23646 & 01 & 04 & 38.88 & +02 & 01 & 18.18 &  21.61 &   0.08 &  21.03 &   0.20 &  21.22 &   0.31\\
  15705 & 01 & 04 & 42.84 & +02 & 01 & 18.40 &  19.90 &   0.02 &  19.37 &   0.03 &  19.24 &   0.05\\
  24995 & 01 & 04 & 43.08 & +02 & 01 & 18.50 &  21.96 &   0.11 &  21.23 &   0.14 &  20.69 &   0.28\\
  07510 & 01 & 04 & 51.93 & +02 & 01 & 18.54 &  19.43 &   0.01 &  18.70 &   0.02 &  18.57 &   0.03\\
  24468 & 01 & 04 & 58.15 & +02 & 01 & 18.83 &  21.15 &   0.06 &  20.36 &   0.09 &  20.31 &   0.12\\
  17323 & 01 & 04 & 40.09 & +02 & 01 & 18.97 &  20.30 &   0.03 &  19.70 &   0.04 &  19.64 &   0.07\\
  02009 & 01 & 05 &  5.71 & +02 & 01 & 19.22 &  20.71 &   0.06 &  19.94 &   0.08 &  19.07 &   0.06\\
  25222 & 01 & 04 & 57.83 & +02 & 01 & 19.44 &  21.06 &   0.05 &  20.50 &   0.08 &  20.48 &   0.14\\
\enddata 
\tablecomments{Sample photometry for 10 stars of the full catalog, the remainder of which can be accessed through the online journal.} 
\end{deluxetable*} 

To generally avoid contamination from evolved intermediate-aged stars in the main disk of \ic,  we selected only those stars outside of the main body of the galaxy, as shown in \autoref{fig:IC1613xy}. The boundary for the outer disk sample was chosen such that 
the majority of the stars seen in the CMD are Population II.
Out of the 23,525 stars in our catalog, approximately half (12,358 stars) are in our halo sample. We further filtered sources for a signal-to-noise of $\geq10$, resulting in 6,070 high signal-to-noise ratio stars. The CMD of this high signal-to-noise, predominantly lower-density population of stars is given in \autoref{fig:IC1613-JJK-CMD-HALO-4-17-mark}. \autoref{tbl:tbl1} contains the positions, $JHK$ magnitudes and errors for a sample of 10 stars from the full catalog, the remainder of which can be found in the online journal.

\section{Identifying the Tip of \\the Red Giant Branch}

\begin{figure*} 
\centering 
\includegraphics[width=12cm,angle=-90]{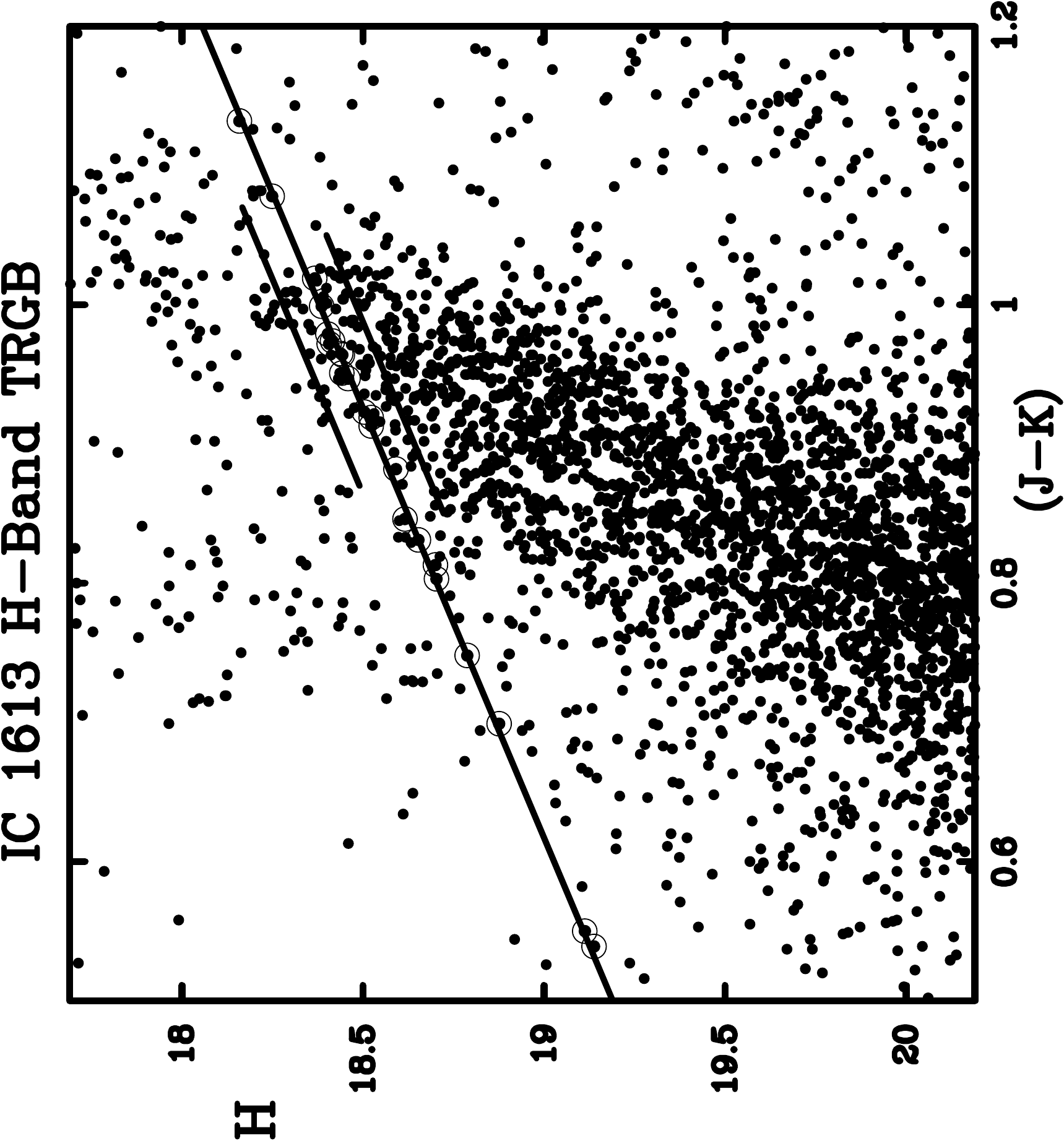} \caption{\small The $H$ vs $(J-K)$ color-magnitude diagram for Population~II stars in the halo of \ic. The solid line shows the adopted slope and position of the TRGB sequence as a function of the
metallicity-sensitivity of the $H$-band magnitude to the $(J-K)$ color. Circled dots are $\pm$0.02 mag above and below the line defining the TRGB. They are also individually cross-identified in each of the two panels in \autoref{fig:IC1613-IR-TRGB-raw-JJK-KJK-4-17-mark}. In moving from $H$ to $J$ and $K$ their independent photometric errors  inflate the $H$-band defined TRGB stars away from the fiducial line. The two flanking lines covering the color range of the TRGB are arbitrarily displaced in magnitude by $\pm0.10$~mag so as to visually highlight the scatter in the color range of the TRGB.
\label{fig:IC1613-IR-TRGB-raw-HJK-4-17-mark}} 
\end{figure*}

As described earlier and displayed in \autoref{fig:TRGB-diag}, it is predicted that the stars defining the end of the (hydrogen shell-burning) evolution of first-ascent RGB stars will individually trace differently sloping lines  of peak luminosity. These lines are a function of color (as driven by the metal content of stellar atmospheres). As is well known in the $I$-band, the stars defining the tip of the RGB exhibit a shallow slope\footnote{\citet{mad09} found a downward trend of $\Delta I / \Delta (V-I) = $+0.20$\pm$0.05~mag/mag in NGC~4258; \citet{riz07} earlier found essentially the same value 0.22$\pm$0.02 using a larger sample of galaxies, resulting in a higher quoted precision in the slope.} 
as a function of color (any color). Longward (or shortward) of the $I$-band in wavelength, the slope of the termination points of the RGB must be positive (or negative) as a function of color. The longward prediction is affirmed in and illustrated by Figures \ref{fig:IC1613-JJK-CMD-2-17} and \ref{fig:IC1613-JJK-CMD-HALO-4-17-mark}.

\begin{figure*} 
\centering 
\includegraphics[width=12.0cm,angle=-90]{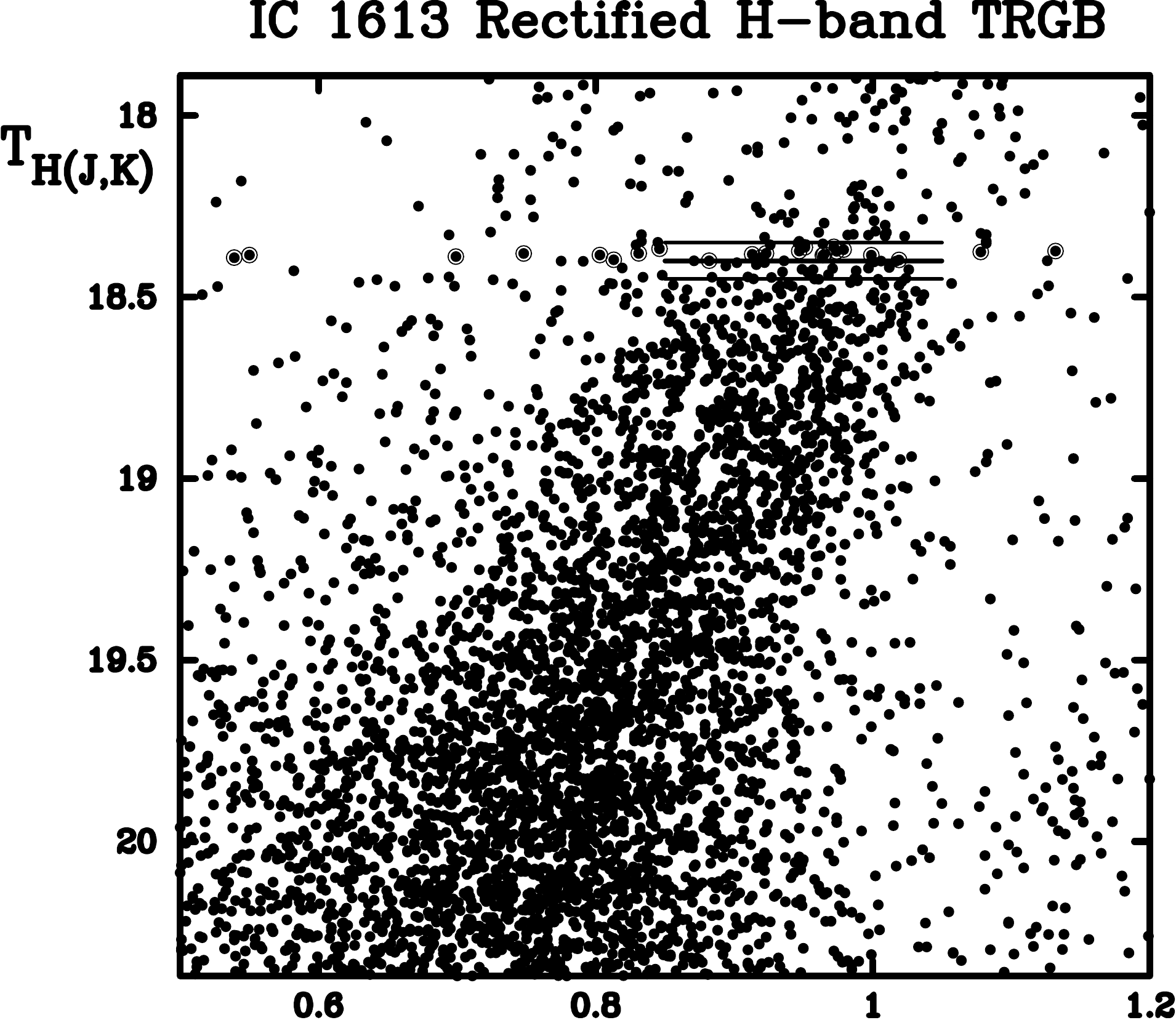} 
\caption{\small The rectified $H$ vs $(J-K)$ color-magnitude diagram for stars in the halo of \ic. All of the data and the TRGB fiducial line have had an affine transformation applied to them such that the TRGB
run with color is neutralized. It is in this projection that the marginalized luminosity
function, shown in Fig. 10 was made. Three short, horizontal lines mark the color width
over which the TRGB was measured. The magnitude separation of the outer two lines is reset
to $\pm$0.05 mag. Stars within $\pm$0.02~mag of the central line are shown as circled dots
highlighting the tip and tracing its formal extrapolation across the CMD. These same
defining stars are re-identified in each of the other CMDs.
\label{fig:IC1613-IR-TRGB-HJKrect-4-17-mark}} 
\end{figure*}

In \autoref{fig:IC1613-IR-TRGB-raw-HJK-4-17-mark}, we see a zoomed-in view of the \ic outer-disk CMD in $H$ vs $(J-K)$, focusing on the upper two magnitudes of the RGB. The upward-sloping TRGB is marked in two ways: first, there is a solid straight line slanting across the CMD having the slope of the TRGB trace
(see description of the methodology below); and second, stars that fall within $\pm0.02$~mag of the fiducial TRGB line given above are shown as circled dots. These stars define the TRGB in the color range $0.85 < (J-K) < 1.05$~mag 
(i.e., metallicity range) found in \ic old Population~II stars. Circled stars outside this color range are also shown so as to better delineate a transformation (discussed later in the text) of this line into the $J$ and $K$ vs $(J-K)$ CMDs.

\begin{figure*} 
\centering 
\includegraphics[width=16.0cm, angle=-90]
{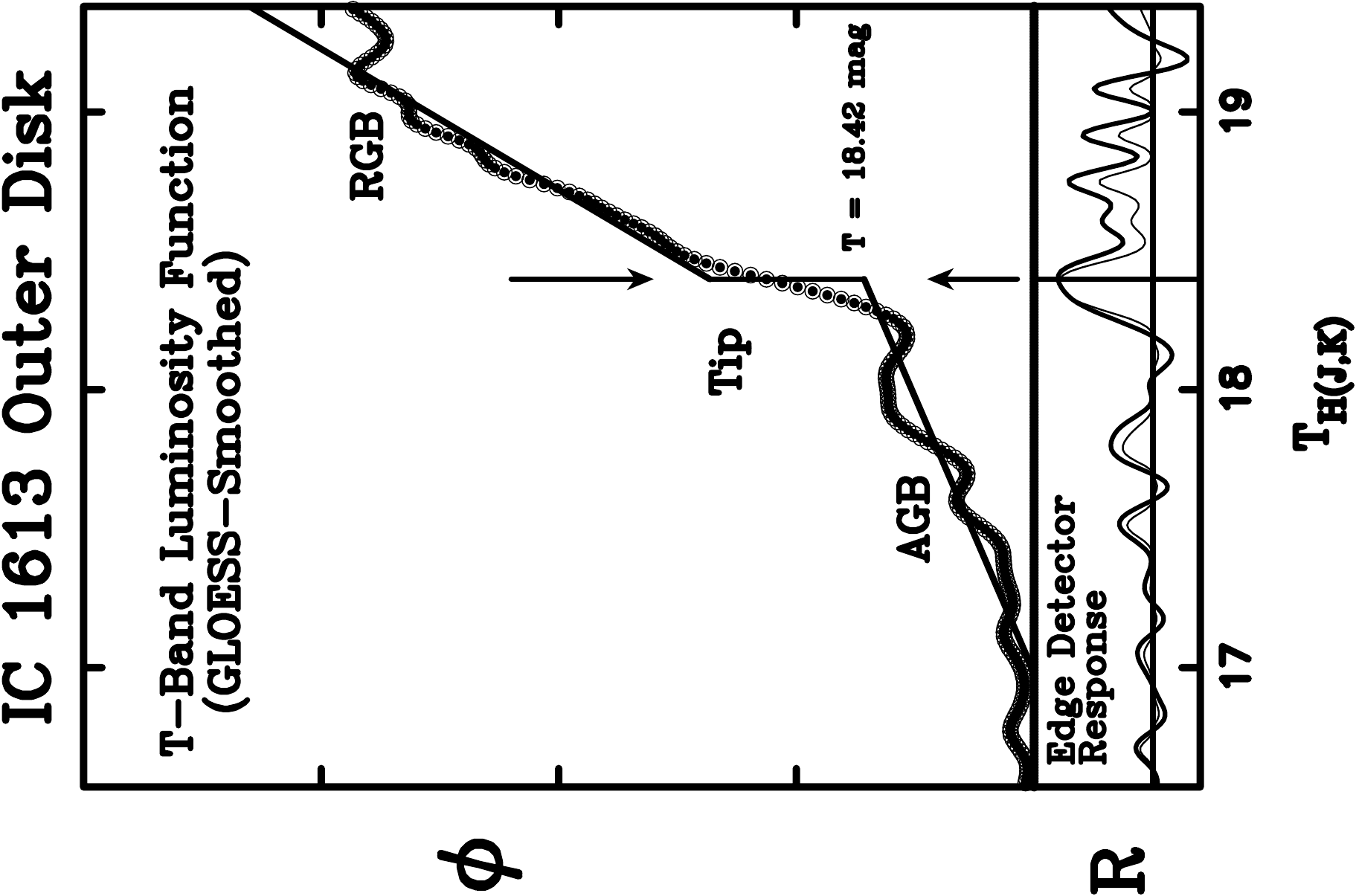} 
\caption{The near-infrared luminosity function
and TRGB detection for Population~II stars in \ic. Straight lines marking the AGB, RGB
and tip discontinuity are shown for illustrative purposes only; they are not used anywhere in the analysis.
\label{fig:IC1613-IR-TRGB-HJK-2-17-mark}} 
\end{figure*}

The fiducial lines in \autoref{fig:IC1613-IR-TRGB-raw-HJK-4-17-mark} were produced in the following self-consistent way: we visually made a first approximation of the slope and zero-point of the TRGB locus in the $H$~vs~$(J-K)$ CMD. Using that slope the data were then transformed into the $T[H,(J-K)]$ plane as first introduced and implemented in \citet{mad09}. The transformed T-magnitude CMD is given in \autoref{fig:IC1613-IR-TRGB-HJKrect-4-17-mark}.
We then constructed the T-band luminosity function by marginalizing over color in the T-band CMD. The marginalized T-band luminosity function was constructed having bins of width equal to 0.01~mag. The GLOESS-smoothed\footnote{GLOESS stands for Gaussian-windowed, Locally-Weighted Scatterplot Smoothing, a modern variant on a method first developed by \citet{Clev1979}. We first used Gaussian windowing for Cepheid light-curve smoothing in \citet{per04}, subsequently in \citet{sco11,sco16} and most recently in \citet{mon17} for RR Lyrae light-curve smoothing where additional details and descriptions are given. Most recently, \citet{hatt17} have discussed the application of GLOESS to luminosity-function smoothing and TRGB detection.} edge detector was iteratively applied to the luminosity function where the slope of the TRGB locus was perturbed around the initial approximation until a minimum width in the tip detection output was achieved.

This solution is shown graphically in \autoref{fig:IC1613-IR-TRGB-HJK-2-17-mark} where the tip is detected at 20.01~mag. The filter response at that magnitude has a width of $\sigma = 0.065$ mag. With 125 stars contributing to the response at that magnitude, the formal error on the TRGB magnitude is $\pm 0.006$~mag. We conservatively adopt $\pm 0.01$~mag as the statistical uncertainty in measuring the peak. 
The long solid lines in Figures \ref{fig:IC1613-IR-TRGB-raw-HJK-4-17-mark} and \ref{fig:IC1613-IR-TRGB-HJKrect-4-17-mark} correspond to the adopted $H$-band TRGB solution discussed above.

\begin{figure*} 
\centering
\includegraphics[width=7.5cm,angle=-90]{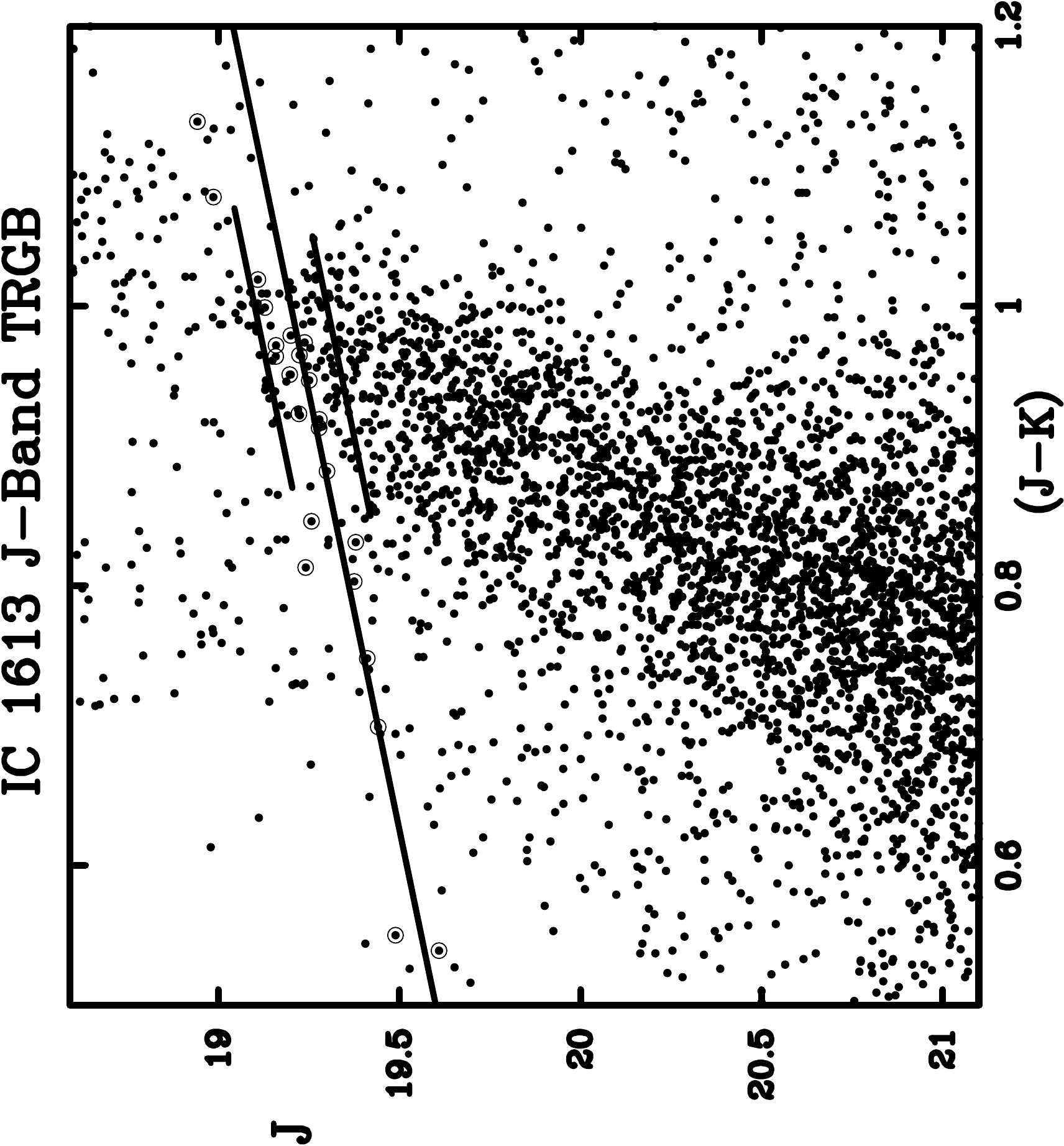}
\vspace{0.25cm} 
\includegraphics[width=7.5cm,angle=-90]{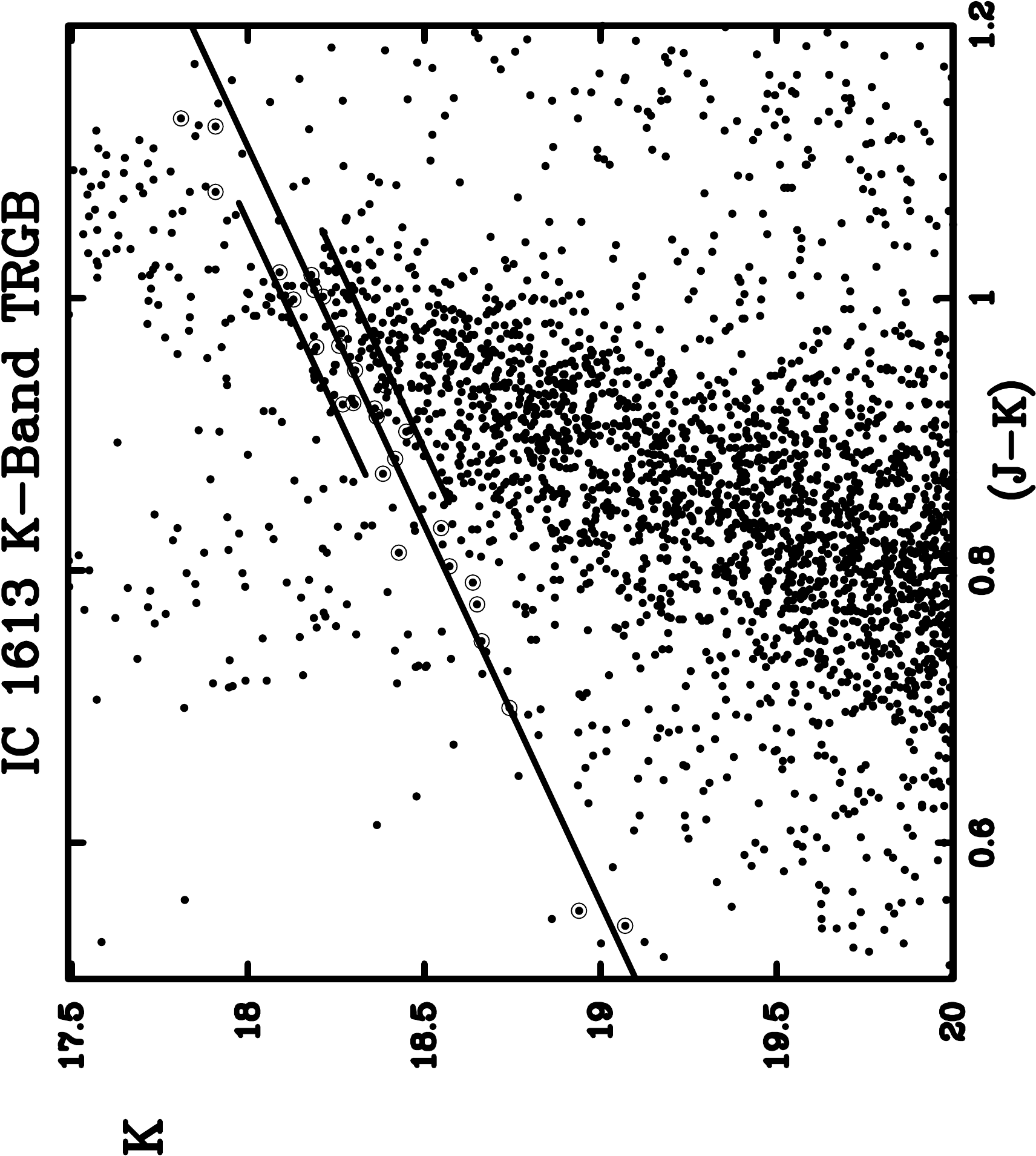}
\includegraphics[width=7.5cm,angle=-90]{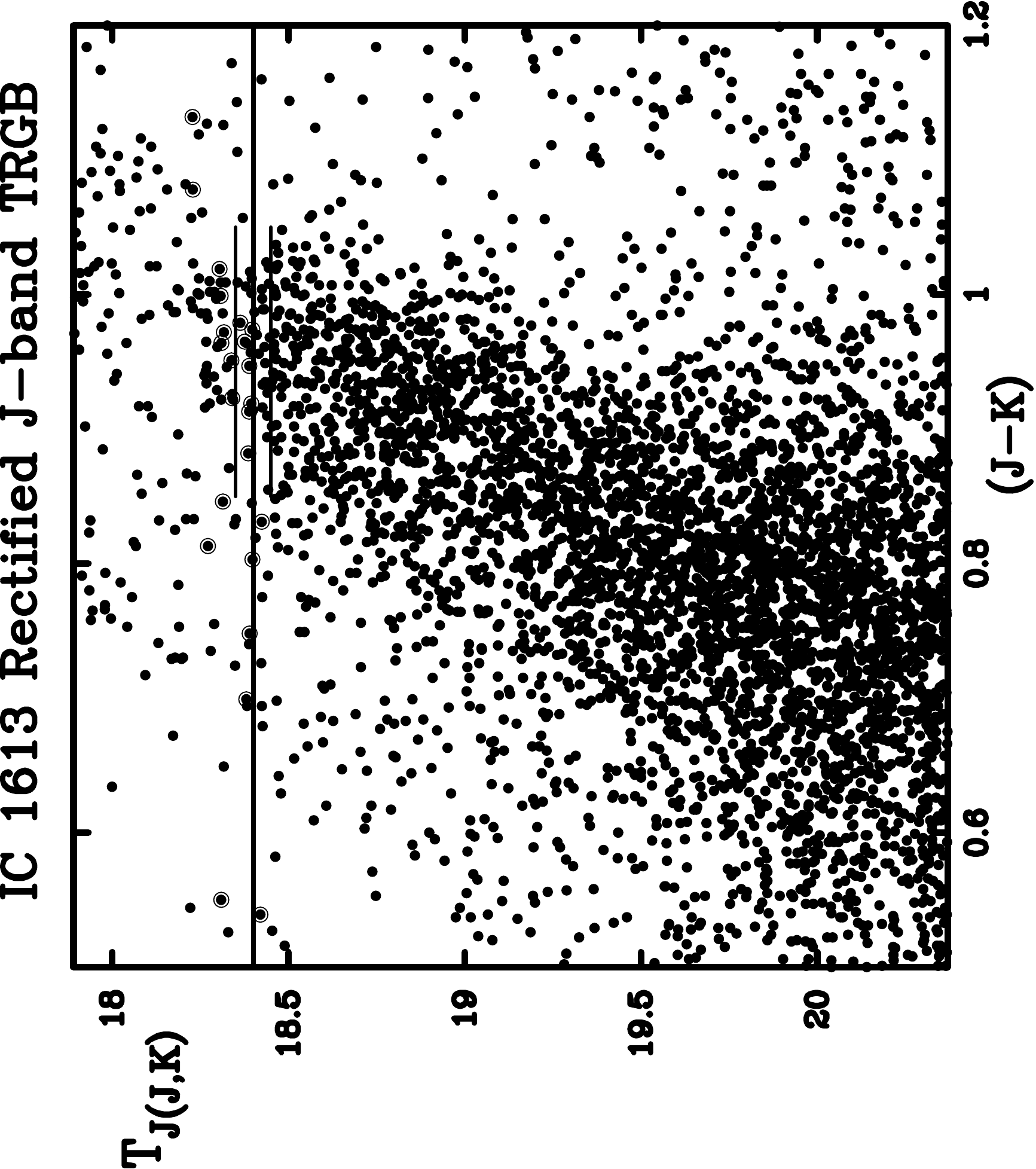} 
\includegraphics[width=7.5cm,angle=-90]{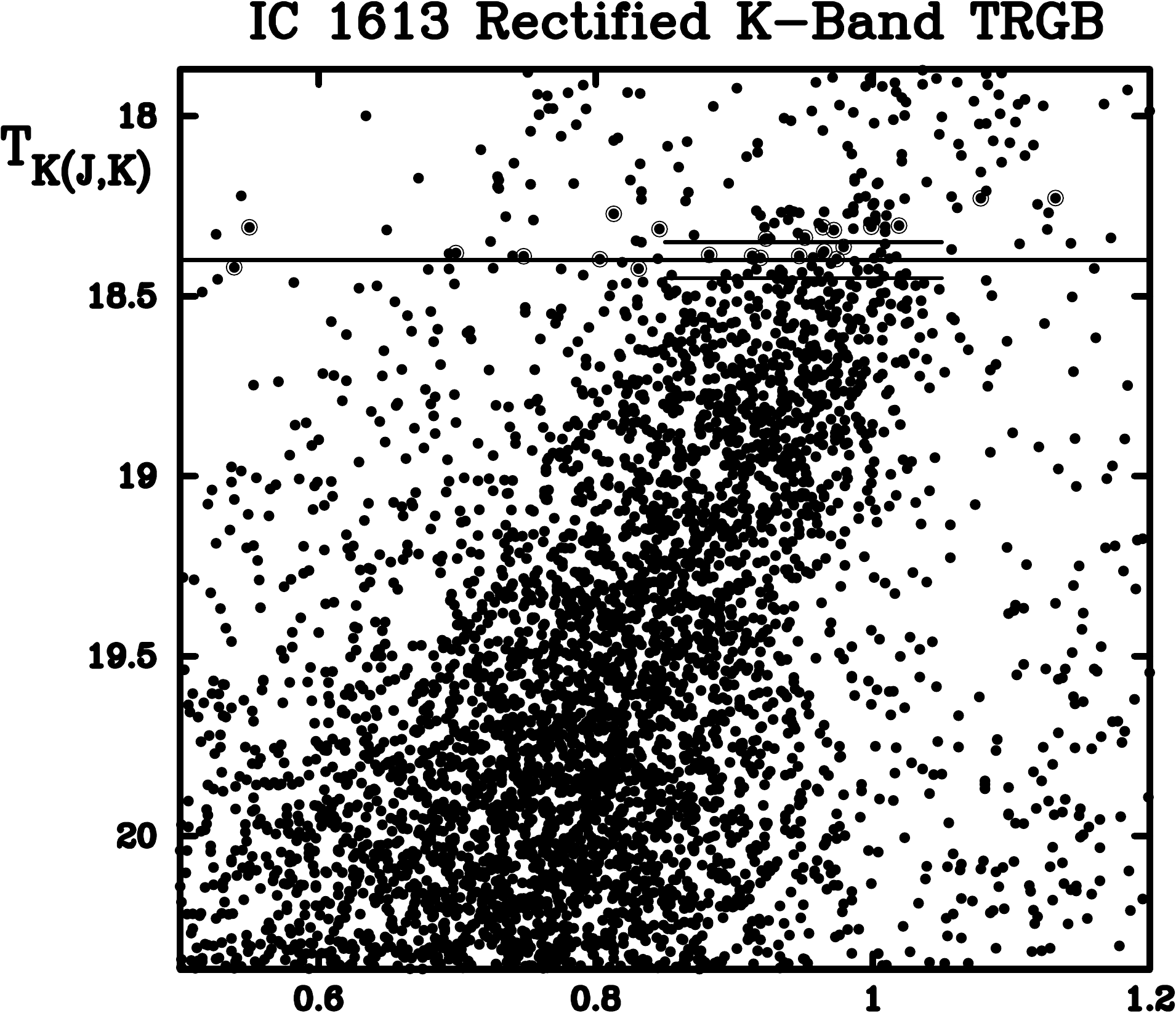}
\caption{(a): The $J$ vs $(J-K)$ color-magnitude diagram for predominantly Population~II
stars in the halo of \ic; (b): Same as (a) for $K$ vs $(J-K)$; (c) and (d): Same as \autoref{fig:IC1613-IR-TRGB-HJKrect-4-17-mark} for T$[J,(J-K)]$ and T$[K,(J-K)]$.
The central solid line in each plot shows the adopted slope and position of the TRGB sequence as a function of the metallicity-sensitivity.
The circled data points are identically the same stars as chosen in \autoref{fig:IC1613-IR-TRGB-raw-HJK-4-17-mark}, but now more widely dispersed because of their independent photometric errors in these plots. 
The short solid lines flanking the fits are drawn over the color range occupied by the TRGB stars, displaced in magnitude simply to represent the photometric scatter in the stars selected to define the tip population in these plots. }
\label{fig:IC1613-IR-TRGB-raw-JJK-KJK-4-17-mark}
\end{figure*}

In panels (a) and (b) of \autoref{fig:IC1613-IR-TRGB-raw-JJK-KJK-4-17-mark} we show the CMDs for the $J$-band and for the longer-wavelength $K$-band, respectively, holding the $(J-K)$ colors in common across the CMDs. The solid lines in these two cases correspond to the self-consistent mapping of the $H$-band solution, as well as the stars marked in the $H$ vs $(J-K)$ CMD mapped into these $J$ and $K$ CMD planes. This makes mathematical\footnote{Translations, rotations, scalings, reflections, shear,  dilations and compressions all fall under the umbrella of affine transformations. Affine transformations all have one thing in common: before and after an affine transformation collinear points remain collinear, and parallel lines remain parallel. The lines and points in Figures \ref{fig:IC1613-IR-TRGB-raw-HJK-4-17-mark} \& \ref{fig:IC1613-IR-TRGB-raw-JJK-KJK-4-17-mark}, for example, undergo an affine rotation and so preserve their relative ordering when they are seen in Figures \ref{fig:IC1613-IR-TRGB-HJKrect-4-17-mark} \& \ref{fig:IC1613-IR-TRGB-raw-JJK-KJK-4-17-mark}.}  and physical sense; however, exceptions (generated by noise) will arise for individual stars, when photometric measurement uncertainties scatter stars differentially around the same line seen from bandpass to bandpass. The corresponding T-magnitude plots for these CMDs are given in panels (c) and (d) of \autoref{fig:IC1613-IR-TRGB-raw-JJK-KJK-4-17-mark}.

As stated in Section \ref{sec:data}, the result above is based on the outer-disk sample of stars. The question naturally arises as to whether there is any detectable difference in the detected tip magnitude between the global sample and the more restricted outer-disk sample. The simple answer is no; we detect no statistically significant difference between the inner and outer regions of \ic. Nevertheless, as this paper was being finalized we endeavored to obtain new \emph{FourStar} imaging further out into the actual halo of \ic. Those observations were made, but they were taken under poorer seeing conditions and the signal-to-noise at the tip was significantly degraded with respect to the first dataset being discussed here. However, the new data were sufficient to show that at the $\pm0.02$~mag level, there is still no gradient in the apparent magnitude of the TRGB in going from the central regions out into the extended halo of this galaxy. Independent confirmation of this lack of any radial gradient in the TRGB is also found in the recent literature. \citet{sib14}, for example, obtained high-quality $JHK$ stellar photometry over a 0.8 square degree region centered on \ic. They studied both the azimuthal and radial dependence of the TRGB magnitude and concluded that ``there is no overall trend'' in either case (specifically their Figure 12 shows no gradient out to 4~kpc). However, we remain focused on the outer-disk component of the data, which is more than substantial enough in star count for a high-fidelity measurement of the TRGB and its structure, in order to ensure a clean and convincing calibration result.
\clearpage
\section{Relative Zero-Point Calibrations}
\label{sec:relative-zero-point-calibrations}

Undertaking tip detection independently for each of the slope-corrected $JHK$ luminosity functions give the following apparent magnitude
relations for the TRGB in \ic:
\begin{align}
J &= 19.21 - 0.85 \left[(J-K) - 1.00\right]\\
H &= 18.40 - 1.62 \left[(J-K) - 1.00\right]\\
K &= 18.21 - 1.85 \left[(J-K) - 1.00\right].
\end{align}

\noindent Errors on the slopes are $\pm$ 0.12, 0.22 and 0.27 in $J,H$ \& $K$,
respectively, and the equivalent calibrations in terms of $(J-H)$ colors are:
\begin{align}
J &= 19.24 - 1.11 \left[(J-H) - 0.80\right]\\
H &= 18.44 - 2.11 \left[(J-H) - 0.80\right]\\
K &= 18.26 - 2.41 \left[(J-H) - 0.80\right].
\end{align}

\par\noindent Errors on these slopes are $\pm$ 0.15, 0.26 and 0.36 in $J,H$ \& $K$,
respectively. Random errors on the zero-points are $\pm$0.03 mag for all three bands. We discuss the color-color relation in the following Section as well as the Appendix and only report the result for the \ic TRGB here:
\begin{align}
(J-H) = 0.77 \left(J-K\right) + 0.04.
\end{align}

In a companion paper on the NIR-TRGB calibration in the LMC (Hoyt et al.,  2018 submitted) we derive an absolute calibration of the NIR-TRGB method using stars in the bar that are co-extensive with the detached eclipsing binaries geometric distance moduli that give a mean of 18.49 $\pm$0.01 (statistical) $\pm$0.05 (systematic)~mag \citep{pie13}. In Hoyt et al. we give the following absolute calibrations:
\begin{align}
M_J &= -5.14 - 0.85 \left[(J-K)_o - 1.00\right]\\
M_H &= -5.94 - 1.62 \left[(J-K)_o - 1.00\right]\\
M_K &= -6.14 - 1.85 \left[(J-K)_o - 1.00\right], 
\end{align}
\noindent
and 
\begin{align}
M_J &= -5.13 - 1.11 \left[(J-H)_o - 0.80\right]\\  
M_H &= -5.93 - 2.11 \left[(J-H)_o - 0.80\right]\\
M_K &= -6.13 - 2.41 \left[(J-H)_o - 0.80\right].
\end{align}

Applying these LMC calibrations to the \ic data, and adding in quadrature the LMC statistical and systematic error estimates, gives an average distance modulus of
24.32 $\pm$0.02 (statistical) $\pm$0.06~mag (systematic), which is in excellent agreement with the distance modulus of 24.30~mag and its uncertainty ($\pm0.05$~mag) adopted by \citet{hatt17}. Furthermore, our derived value of 24.32~mag is, to within the errors, the same as the multi-wavelength Cepheid distance modulus of 24.29 $\pm$ 0.03~mag, most recently determined by \citet{sco13}, which in turn is also equal to the median Cepheid distance modulus of 24.31~mag, based on 51 published Cepheid distances found in NED-D. Finally, our value agrees with the median distance modulus of 24.30~mag determined from 14 RR~Lyrae distance moduli listed in NED.

As discussed in Madore \& Freedman (2018 in prep.)  
the three ($JHK$) solutions for the relations 
of TRGB magnitude with color are tightly coupled mathematically. For instance, once the color/color relation and one of the color/tip-magnitude relations are adopted then all of the other color/magnitude combinations are mathematically determined. In the Appendix we discuss this particular path to a calibration as an external consistency check on the more traditional approach given above.

\subsection{Reddening Considerations}
Using multi-wavelength observations of Cepheids, \citet{sco13} determined a total line-of-sight reddening $E(B-V) = 0.05$~mag into the main body of \ic. Adopting a ratio of total-selective absorption of $R_V = $ 3.1 and conversion factors of 0.282, 0.190 and 0.114 for mapping $A_V$ values into $JHK$ extinctions \citep{card89} the main-body extinctions would be $A_J = $0.044 $A_H = $0.033 and  $A_K = $0.018~mag. However, our stars are in the outer disk of \ic, which is almost certainly devoid of any appreciable amounts of dust. 
NED quotes a Galactic forefround reddening of $E(B-V) = $ 0.025~mag. In addition, \citet{san71} used photoelectric $UBV$ photometry of galactic stars across the face of \ic to evaluate the foreground reddening. In his own words he states ``The data {\it require $E(B-V) \le 0.03$, and it could be zero.} In any case, it is almost certainly smaller than $E(B-V) = 0.07$ ..." (emphasis ours). Taking each of these points into consideration we adopt a foreground correction of $E(B-V) = $ 0.03~mag and propagate it into the derivation of the three true distance moduli, averaged, and given above.

\clearpage-crop-crop.pdf
\section{Comparisons with Theory} 

\begin{figure*} 
\centering 
\includegraphics[width=14.0cm,
angle=-90]{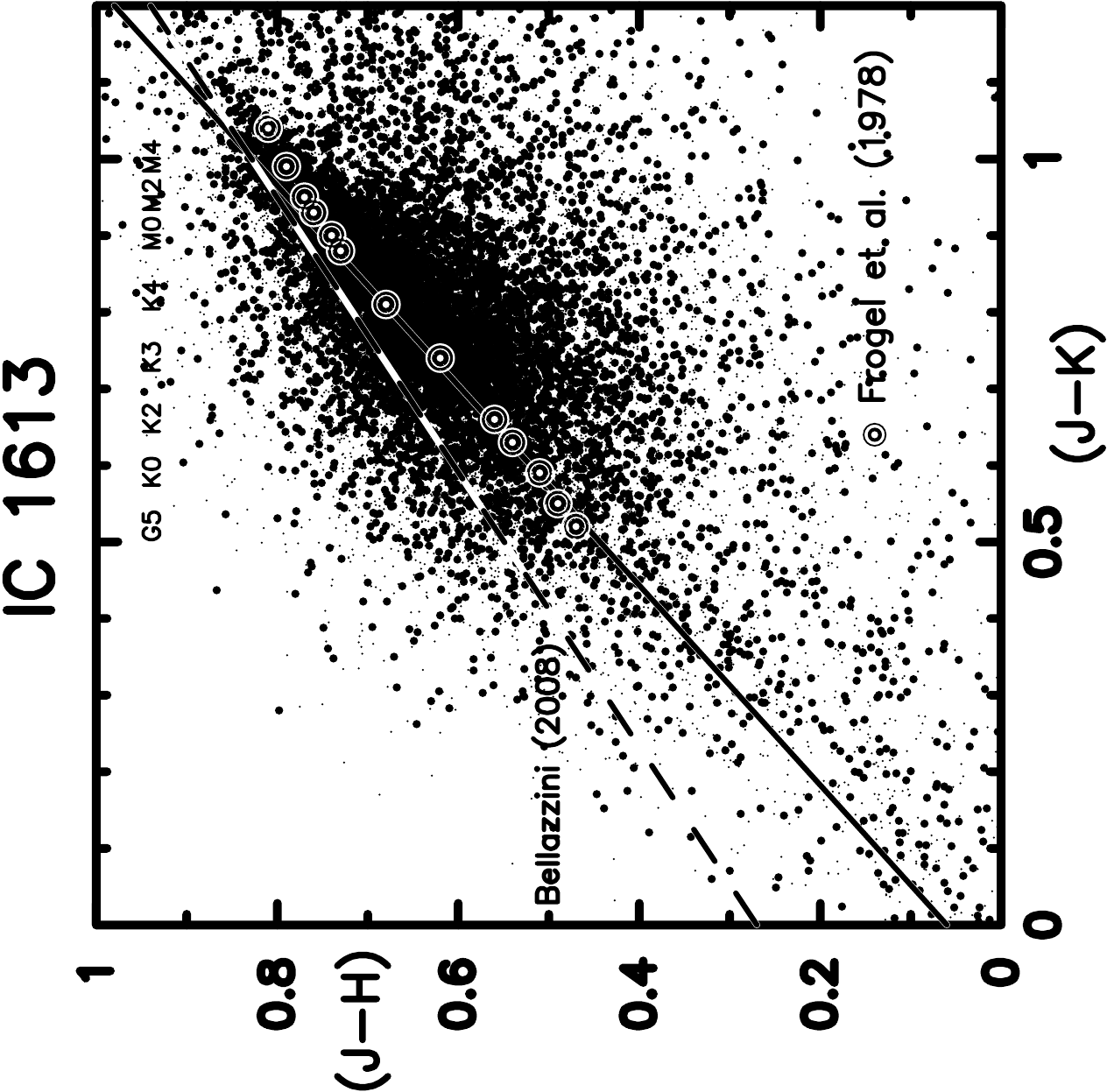} 
\caption{$(J-H)$ vs $(J-K)$ color-color plot for stars in \ic. Large dots are stars in the halo of IC 1613; small dots are in the main body. Large circled dots are at the positions of the adopted mean colors for Milky Way giants taken from \citet{fro78}, their Table 12. The solid black line is our adopted fit to the \citet{fro78} data for the K0 to K5 giants, appropriate to the TRGB. The run of spectral types with $(J-K)$ color are shown across the upper right portion of the figure. Finally, the theoretical run of TRGB colors taken from \citet{bel08} are shown as the broken line displaced above and having a shallower slope than the empirical (Galactic and \ic) data.
\label{fig:IC1613color-color-a-4}} 
\end{figure*}

Based on fitting isochrone models from \citet{bre12}, \citet{wu14} present a calibration of the TRGB luminosity (in the F110W and F160W \emph{HST} flight magnitudes which closely match $J$ and $H$ ground-based magnitudes) as a function of color. In the color range $(J-H) = 0.70$ to 0.95~mag they find $M_J = -4.81$~mag and $M_H = -5.61$~mag at our fiducial color of $(J-H)_0 = 0.80$~mag. These are to be compared with our values of  $M_J = -5.10 \pm 0.03$~mag and $M_H = -5.90 \pm 0.03$~mag, with the differences in both bands being +0.29~mag, in the sense that we are brighter than the \citet{wu14} values. Our respective slopes agree at the 12\% level over the quoted color range, with the \citet{wu14} slopes being steeper than ours.

Using different published theoretical models \citep[primarily the BASTI set from][]{pie04,cor07}, \citet{bel08} gives color-dependent calibrations of the TRGB using the $(J-K)$ color. Here we give his calibrations (which, it should be noted are offset from the original models by +0.24~mag, so as to fit his globular cluster distances), re-centered on $(J-K)$ = 1.00~mag:
$M_J = -5.37 - 1.08 [(J-K)_0 - 1.00]$,
$M_H = -6.20 - 1.64 [(J-K)_0 - 1.00]$,
and $M_K = -6.37 - 2.08 [(J-K)_0 - 1.00]$. Across all bands, the \citet{bel08} calibration is found to be -0.22~mag
brighter than our adopted calibration via \ic. In addition, his slopes are all
consistently higher (by about 10\%) than the slopes derived in our study. This current
offset between zero-points is within the uncertainties allowed by the already comparably
large zero-point offsets applied by \citet{bel08} to get correspondence between theory and
the observations available to them at their time of writing.

These two independent comparisons of our calibrations with theory find no consensus, even as to the sign of the difference. Put into stark contrast, the two theoretically-based TRGB calibrations in the NIR differ from each other by over 0.5~mag. We take this as a further indication of the real need for establishing a calibration empirically.

There are further implications regarding the color-color relation between \citet{bel08} and our result (given in Section \ref{sec:relative-zero-point-calibrations} and further discussed in the Appendix). Pairwise differencing the \citet{bel08} $J$ and $H$ vs $(J-K)$ equations gives rise to his version of the color-color $(J-H)$ vs $(J-K)$ relation, viz, 
$(J-H)_0 = 0.56\left(J-K\right)_0 + 0.27.$
\noindent This slope (0.56) is shallower and the intercept (0.27 mag) is redder in $(J-H)$
than ours (0.77 and 0.06 mag,respectively). This comparison is more clearly seen in \autoref{fig:IC1613color-color-a-4}, where we have plotted the \citet{bel08} calibration as a slanting broken line, and our
empirical calibration as the solid line. This theory-based calibration is at variance with
the \citet{fro78} individual data points, and divergent both with the fit to the
Milky Way data (the solid line) and with the \ic stellar data being presented here.
These current results and those forthcoming by \emph{Gaia} will offer increasingly accurate
constraints on the theoretical calibration of the TRGB.

\begin{figure*} 
\centering 
\includegraphics[width=\textwidth,angle=0]{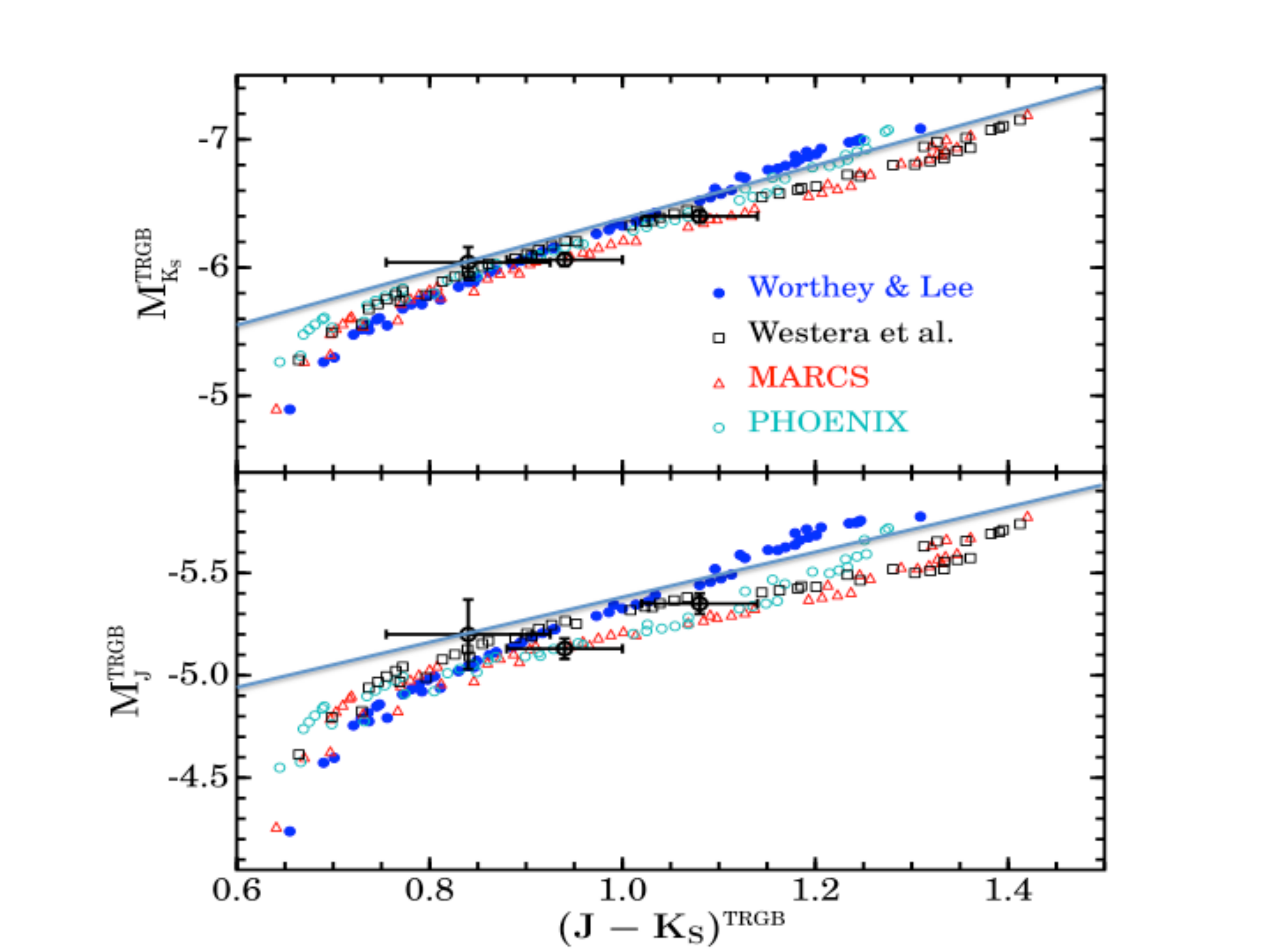} 
\caption{Magnitude of the TRGB in the K band (upper panel) and the J band (lower panel) as a function of (J-K) color as taken from \citet{ser17}. The various colored points are models of the TRGB as noted in the upper panel. The solid line is our empirical calibration of the slope and absolute zero point of TRGB. Our slope is in good agreement with the MARCS, PHOENIX and Westera models \citep[see references within][]{ser17}, while our zero point is systematically brighter.
\label{fig:Slide1}} 
\end{figure*} 

Finally, we note that as this paper was coming to completion, \citet{ser17} published an extensive compilation and update of theoretical models of the TRGB covering a wide range of stellar masses and metallicities, defining the bolometric magnitude at the tip as a function of effective temperature. Applying a variety of modern bolometric corrections then allowed them to map these models into commonly adopted photometric band passes, specifically $VI$ in the optical and, most importantly for our purposes, $JK$ in the near-infrared. \autoref{fig:Slide1} shows an adaptation of their Figure 11 over-plotted with our $J$ and $K$ vs $(J-K)$ calibrations. 
The correspondence is gratifying and clearly favors the slope of the MARCS and PHOENIX bolometric corrections, as discussed in their paper, over the other two contenders 
within the narrow range of colors ($0.85 < (J-K) < 1.05$) available to us for the \ic calibration. The difference between the \citet{ser17} zero-point and ours (pivoting at $(J-K) = 1.00$) is about 0.1~mag, with the theoretical value being fainter. This difference is well within the 0.25-0.30~mag full range of TRGB tip magnitude systematic uncertainty quoted by \citet{ser17} in their Section 5.1. This difference might now be inverted to put interesting constraints on the stellar interior input physics.

\subsection{Multi-Wavelength Trends and Correlations}

\begin{figure*} 
\centering 
\includegraphics[width=12.0cm, angle=-90]{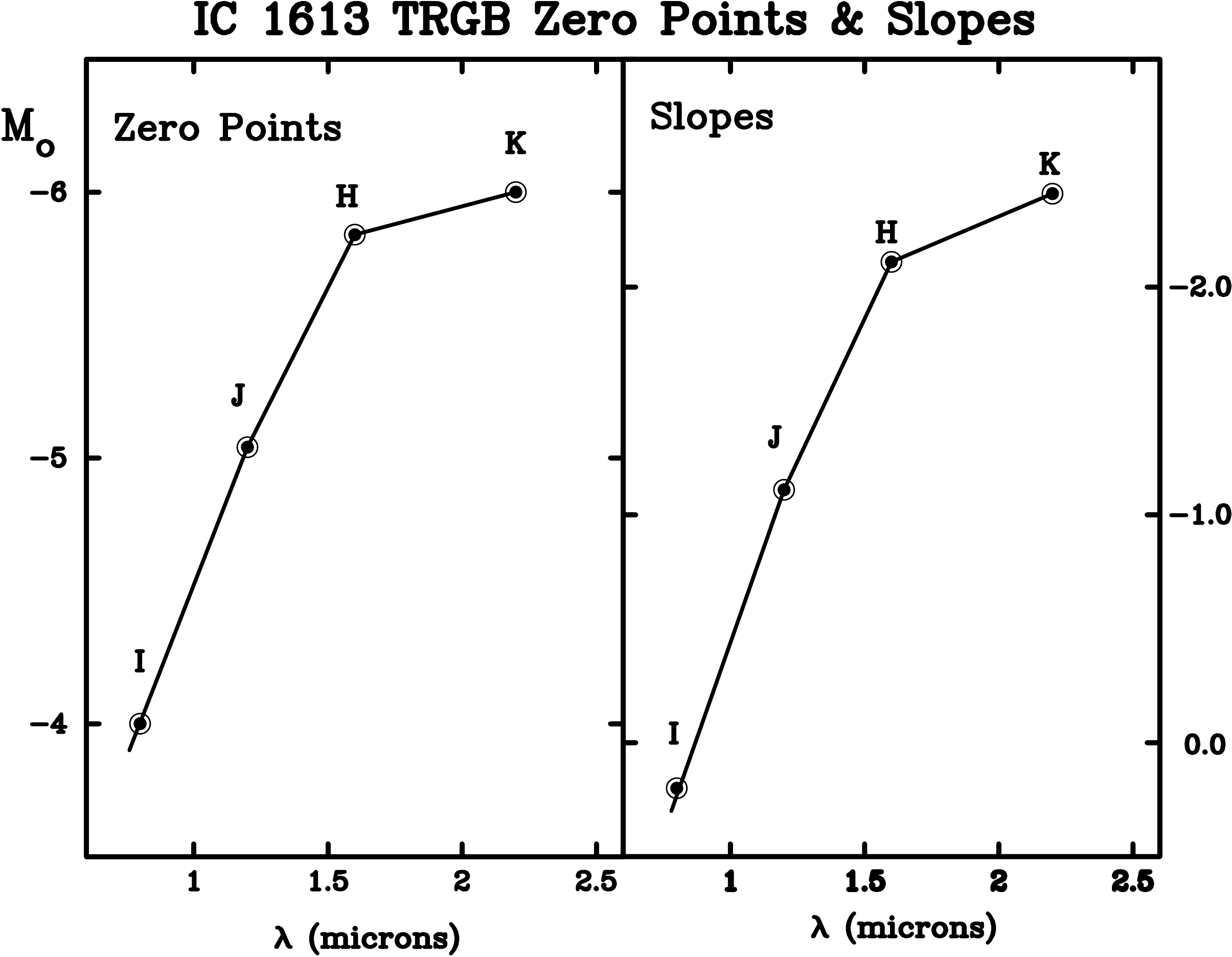}
\caption{\small Zero-point and slope of the TRGB with $(J-H)$ color as a function of bandpass. Slopes are plotted in the right panel. Zero-points, measured at $(J-H) = 0.80$~mag, are plotted in the left panel. The general trend is for the characteristic absolute magnitude of the trace of the TRGB with color to become brighter with increasing wavelength and for the slope to increase as well. Both quantities appear to be starting to level off around 2 $\mu$m. 
\label{fig:IC1613-slope-zp-2}} 
\end{figure*}

\begin{figure*} 
\centering 
\includegraphics[width=12.0cm, angle=-90]{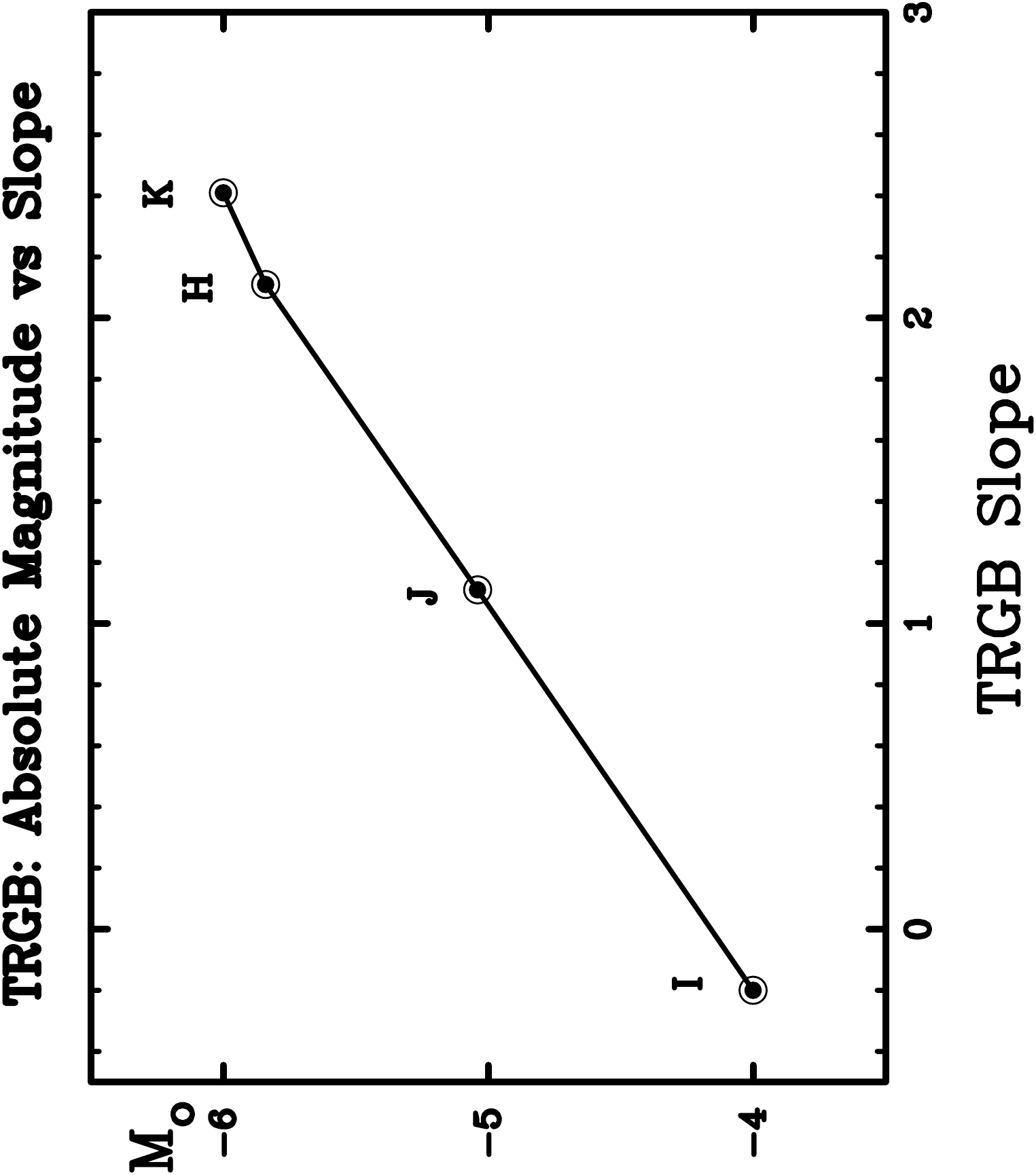} 
\caption{\small Absolute magnitude of the TRGB as a
function of TRGB slope for $IJH~ \& ~K$ bandpasses.
\label{fig:IC1613-slope-vs-zp-2}} 
\end{figure*}

In \autoref{fig:IC1613-slope-zp-2} we show the  multi-wavelength $(IJHK)$ relation of the absolute magnitude zero-point of the TRGB (left panel) and its slope (right panel). The solid lines joining the data points suggest that both functions are monotonically increasing, but slowing in their mutual growth as a function of wavelength passing from the near to the mid-infrared. \autoref{fig:IC1613-slope-vs-zp-2} shows that the slopes and magnitudes are tightly coupled. The origin and explanation of this multi-wavelength property of the TRGB is the subject of discussion in an up-coming paper (Madore \& Freedman, in prep.).

\clearpage
\section{Conclusions}

Using $JHK$ near-infrared photometry of red giant stars in the outer disk of the Local Group
dwarf galaxy, \ic, we have quantified the relationship of the absolute magnitudes of the tip of the
TRGB stars as a function of bandpass and color, and presented a preliminary calibration of the increasing (mean) absolute magnitude of the TRGB as a function of increasing
wavelength. At a fiducial color of $(J-H)$ = 0.80~mag the near-infrared absolute
magnitudes for the TRGB are 1.1, 1.9 and 2.1~mag {\it brighter } at $JH \& K$,
respectively than the $I$-band absolute magnitude of the tip. This means that, at the same signal-to-noise as in the $I$-band, the near-infrared TRGB can be used to probe
cosmological distances that are up to 3/6/7 times larger, corresponding to regions that
are 27/216/343 times more voluminous, respectively, than are currently accessible to
$I$-band TRGB applications{\bf \footnote{These estimates, of course, also depend on the relative sensitivities of the detectors for a given telescope. For $HST$ if, using the on-line Exposure Time Calculators on the HST website, the comparison is made between $I$-band observations using ACS/F814W and $J$-band observations using WFC3-IR/F110W then, the same signal-to-noise can be reached for a distance modulus 1.9~mag larger (i.e., a factor of 2.4 times more distant) for the same integration time, using the WFC3 configuration rather than using ACS.}}.
\medskip

\section*{Acknowledgements} We thank the {\it Carnegie Institution for Science} and the
{\it University of Chicago} for their continuing generous support of our long-term
research into the expansion rate of the Universe. The near-infrared observations discussed
here were taken with the {\it FourStar} camera on the Baade 6.5m telescope at Las
Campanas, Chile. This research has made use of the NASA/IPAC Extragalactic Database (NED)
which is operated by the Jet Propulsion Laboratory, California Institute of Technology,
under contract with the National Aeronautics and Space Administration. Support for this
work was provided by NASA through grant number HST-GO-13691.003-A from the Space Telescope
Science Institute, which is operated by AURA, Inc., under NASA contract NAS 5-26555. MGL
was supported by a grant from the National Research Foundation (NRF) of Korea, funded by
the Korean Government (MSIP) (NRF-2017R1A2B400463). Finally, we thank the anonymous referee for their detailed, helpful and constructive comments.


\appendix

\section*{Color-Color Relations for RGB stars in the Near-Infrared}

Central to the methodology undertaken here, in deriving a self-consistent calibration of the multi-wavelength slopes and zero-points of the TRGB in the near-infrared, is the adoption of an independent calibration of the (distance-independent) color-color relation between the $(J-H)$ and $(J-K)$ intrinsic colors of K and M-type giant stars defining the TRGB. 
One such calibration is implicit in the tabulation of adopted mean colors for Milky Way giants (and dwarfs) given in Table 12 of \citet{fro78}. We have fit for the slope and zero-point of the $(J-H)-(J-K)$ color-color relation using the six fiducial data points spanning K0 to K5, which are representative of the giants defining the TRGB.
We find that
\begin{equation}
(J-H)_0 = 0.77\left(J-K\right)_0 + 0.06  ~\mathrm{(Milky Way)}
\end{equation}

\noindent with uncertainties on the slope and zero-point being $\pm 0.012$ and $\pm 0.008$~mag,
respectively.

This solution and the full complement of giant-star colors from \citet{fro78} are shown in \autoref{fig:IC1613color-color-a-4} overlaid on our sample of halo (large dots) and main body (small points) stars in \ic. Although there is significant scatter in  the \ic data, the overall trend is extremely clear and the fit to the Milky Way giants is seen to be an excellent (and totally independent) representation of the extragalactic sample. However, it must be said that the Galactic sample was not chosen to be representative of the the highest-luminosity RGB stars (i.e., tip stars) alone, and so it may be biased if surface-gravity-dependent color terms are being averaged over in this field sample.

To evaluate this possibility or other indications of non-universality of the TRGB calibration we give the near-infrared color-color relations for TRGB stars in the LMC and
in \ic as discussed in the main text and in Hoyt et al. (submitted).

We find that
\begin{align}
(J-H)_0 &= 0.77\left(J-K\right)_0 + 0.03  ~\mathrm{(LMC)}\\
(J-H)_0 &= 0.77\left(J-K\right)_0 + 0.04  ~\mathrm{(\ic)}
\end{align}

\noindent
where the slopes were fixed. The agreement at the $\pm 0.02$~mag level is encouraging and will be the subject of further scrutiny in future papers.

\clearpage


\end{document}